\documentclass[sn-standardnature]{sn-jnl}% Standard Nature Portfolio 

\usepackage{hyperref}

\jyear{2022}%

\theoremstyle{thmstyleone}%

\theoremstyle{thmstyletwo}%

\theoremstyle{thmstylethree}%

\begin{document}

\title[Ten steps to plan your astronomy paper]{How to plan your astronomy research paper in ten steps}

%%=============================================================%%
%% Prefix	-> \pfx{Dr}
%% GivenName	-> \fnm{Joergen W.}
%% Particle	-> \spfx{van der} -> surname prefix
%% FamilyName	-> \sur{Ploeg}
%% Suffix	-> \sfx{IV}
%% NatureName	-> \tanm{Poet Laureate} -> Title after name
%% Degrees	-> \dgr{MSc, PhD}
%% \author*[1,2]{\pfx{Dr} \fnm{Joergen W.} \spfx{van der} \sur{Ploeg} \sfx{IV} \tanm{Poet Laureate} 
%%                 \dgr{MSc, PhD}}\email{iauthor@gmail.com}
%%=============================================================%%

\author*[1]{\fnm{Nushkia} \sur{Chamba}}\email{nushkia.chamba@astro.su.se}

\author[2,3]{\fnm{Johan H.} \sur{Knapen}}\email{johan.knapen@iac.es}
\equalcont{These authors contributed equally to this work.}

\author[4]{\fnm{Diane} \sur{Black}}\email{d.black2021@outlook.com}
\equalcont{These authors contributed equally to this work.}

\affil*[1]{\orgdiv{The Oskar Klein Centre, Department of Astronomy}, \orgname{Stockholm University}, \orgaddress{\street{AlbaNova}, \city{Stockholm}, \postcode{SE-10691}, \state{}\country{Sweden}}}

\affil[2]{\orgdiv{}, \orgname{Instituto de Astrof\'\i sica de Canarias}, \orgaddress{\street{Calle V\'\i a L\'actea S/N}, \city{La Laguna}, \postcode{E-38205}, \state{}\country{Spain}}}

\affil[3]{\orgdiv{Departamento de Astrof\'\i sica}, \orgname{Universidad de La Laguna}, \orgaddress{\city{La Laguna}, \postcode{E-38200}, \state{}\country{Spain}}}

\affil[4]{\orgdiv{Free-lance English coach}, \orgname{}\orgaddress{\street{}\city{Groningen}, \postcode{}\state{}\country{The Netherlands}}}

%%==================================%%
%% sample for unstructured abstract %%
%%==================================%%

\abstract{Scientific writing is an important skill for a career as a professional astrophysicist. Very few researchers, however, receive any formal training in how to write scientific research papers of high quality in an efficient manner. This paper (Paper I) is the first of a two-part self-help guide in scientific writing to address this skills gap. Paper I focuses on planning your academic research paper in astronomy. We discuss how to crystallise the ideas that underlie a research project, analyse how the paper can be constructed considering the audience and the chosen journal, and give an overview of the publishing process. Paper II is a detailed description of the different sections that make up a research paper in astronomy and shares the best practice in how to write in English. Whether you are a student writing your first paper or an experienced author, you may find the ideas presented here useful.}


%%\pacs[JEL Classification]{D8, H51}

%%\pacs[MSC Classification]{35A01, 65L10, 65L12, 65L20, 65L70}

\maketitle

\section{Introduction}\label{sec1}

Writing papers is key to sharing research and building a career in academia. The top journals in the sciences only accept articles in the English language and a strong publication record is attractive to many employers. Yet most scientists, including astronomers, never receive formal training in academic writing. They may consider that writing is not important or even chose a career in the sciences to avoid a focus on language and writing skills. To complicate matters, many astronomers are non-native speakers of English, so writing a research paper may become a real struggle. \par 

There are numerous books and papers on scientific writing for students or young researchers in biomedical, engineering,  computer science fields \cite[e.g.][]{Ashby2005, Sterk2008, GopenSwan2018, saramaki2018, saramaki2018blog}, including editing, writing and translation services (e.g. \href{http://www.scitext.com/effective-writing/}{Scitext} Cambridge) . These publications discuss the complete writing process, from planning the paper to final editing and polishing, but do not completely meet the needs of astronomy students. Resources that are focussed on astronomy or astrophysics include a series of book chapters detailing the writing process, graphics and ethics \cite{Sterken2011a,Sterken2011b,Sterken2011c}, a handbook aimed at undergraduate and graduate students \cite{Stevance2021}, instructions for authors by journal editors \cite[see e.g.][and references therein]{2004A&A...420E...1B, 2007NatPh...3..581}, and a detailed discussion of how to publish data \cite{Chen2022}. Some publishers also offer writing courses that can be tailored for astrophysics students (e.g. Nature \href{https://masterclasses.nature.com/online-course-in-scientific-writing-and-publishing/16507840}{masterclasses}, \href{https://researcheracademy.elsevier.com/writing-research/fundamentals-manuscript-prepaP2ration}{Elsevier's} course on manuscript preparation or \href{https://academic.oup.com/mnras}{MNRAS's} writing course at the UK National Astronomy Meeting). However, these resources may not be readily available to all students, and academic writing courses are rarely included in the curricula of astronomy and astrophysics in universities around the world. \par

Therefore in this two-part guide we provide concise guidelines in scientific writing for young astronomers worldwide. We target master's or doctoral students but the material may also be of interest to more experienced writers in astronomical research or to writing instructors.
In this paper (Paper I), we describe how to plan your astronomy research paper in ten steps, including how you can begin to put words on paper, and where and how we publish articles in astronomy. In Paper II, we discuss how to write and develop the various sections of a paper in order to collectively build the main message of your research project and share best practice on how to write in correct English, specifically highlighting astronomy-related examples.  \par

\section{The Ten-Step Plan: From Thinking to Paper Submission}\label{sec2}

\subsection{Think Before You Write}
\label{sec_think}
In the experience of NC, writing an abstract for a conference was one of her first scientific writing tasks as a graduate student. However, what she wrote about in her first draft was a summary of the methodology she was developing for her project (as that is what she spent her days working on) rather than the main scientific question her project addressed and what (preliminary) result was achieved thus far in the analysis. She was absolutely surprised when one of her thesis advisors returned a completely different abstract as commentary on her draft, which made her think to herself, `Oh is \textit{that} the whole point my project?!' This is one of many experiences that has led us and other writers \cite[e.g.][]{Sterk2008} to identify the first golden rule on scientific writing: \textit{think before you write}. \par

For instance, to write the Introduction of a paper, you first require knowledge of `\textit{the} literature', i.e. the published body of work in your field (not `literature' which refers to poetry, novels or plays). Actively keeping track of the literature as you work on your project will help you recognise the main sources needed to place your research in the astrophysical context, the gap in the literature that your project aims to fill and how the content in the paper is an important step towards filling this gap (e.g. Is it a new, efficient method? Have you acquired higher-quality data on an astronomical source?). Collecting all these resources to place your project in the broader context of the field is `thinking before writing'. Of course, the thinking needed for each section of the paper will be different and may require feedback from co-authors (consider for example how you think about the interpretation of your results for the Discussion section). \par

When gathering background resources, you also need to read and think critically. Critical reading, thinking and writing are academic practices that many astronomy students may not be familiar with or have never encountered during their education as not all programmes offer such courses. 
To these readers, it is important to understand that to be critical of an idea or paper does not necessarily mean to be negative.  The Oxford Advanced American  \href{https://www.oxfordlearnersdictionaries.com/definition/american_english/critical}{Dictionary} defines critical as `\textit{involving making fair, careful judgements about the good and bad qualities of someone or something}'; in other words, to evaluate the validity of an idea and question when it may or may not hold true. \par 
Obviously, readers can be critical about an idea only if they have sufficient knowledge or expertise about the subject in question. Even more problematic is that many first year doctoral students struggle to read scientific papers. This is not only because the style of scientific writing is very different from novels or popular articles, but very few scientific papers are written well enough for non-experts or students to follow them easily. In other words, it is often not the fault of the reader if a piece of writing seems unclear. \par 

We address these issues with a number of useful tips that new writers in astronomy may find helpful when reading the literature. The guidelines listed below are adapted to astronomy from Wallace \& Wray's book on critical reading and writing \cite[mainly Chapter 4;][]{wallacewray2021}. Following these recommendations may help you skim through papers in an efficient manner to identify which articles could be most relevant to the subject of your research paper and/or take notes as you surf the literature. 

\begin{itemize}

\item Think about why you are reading this paper. Are you looking for an answer to a specific research question? Or evidence to support an assumption used in your study? Write down a list of questions or expectations for your literature review even before you start reading papers.

\item    Why did the authors write this paper? Read the title and abstract first---do they raise more questions than answers to your initial questions? This is completely normal as authors cannot include all the information in a title and abstract. Annotating the abstract as you read it could then help you determine which sections of the paper would be most relevant to further your study.

\item    Study the figures of the paper and read the captions---what story do they tell you? 

\item  What are the main conclusions of the paper? In astronomy, the main claims of the authors can usually be found in the conclusions section. Are these claims warranted by the evidence presented in the text and figures? If not, why not? 

\item  Finally, what use is this paper to you? Go back to point 1 and think about how this paper could be a resource to your own work---does the paper require an in-depth reading? 

\end{itemize}

Navigating the literature with the above questions in mind will help you manage and organise your notes in a meaningful way so that you can come back to them in the future. Note-taking strategies are discussed in for example Sönke Ahrens's \href{https://takesmartnotes.com}{book} where he also points out that new ideas can emerge from notes. So while note-taking does require a time commitment, it might be worthwhile in the long term if the process may lead you to build and create new ideas for future papers, using your own notes! Further discussion of note-taking strategies is beyond the scope of our guide but we encourage readers to explore freely available tools for note-taking, e.g. \href{https://obsidian.md}{Obsidian}.

\subsection{Use Your Thinking Style}
\label{subsec:think_style}

Our golden rule to `think before you write' does not imply that only one thinking style is suitable for scientific writing. Some think primarily in pictures, while others in words \cite{NISHIMURA2015}. Over many years of teaching we have seen that some students need to draw pictures and diagrams to explain an idea---these are visual thinkers. When visual thinkers pick up a new publication, they look first at the figures and graphs. Others tend to talk through problems, and may like to write words and text to think or engage with an idea.  These verbal thinkers pick up a new publication and start reading the text to learn what the paper is about. We have had students who think in terms of graphs, perhaps a variation on thinking in pictures, and others on sounds, colours, or moving processes (the latter are really good at solving a Rubik's cube). And some may immediately see the solution to mathematical problems but cannot easily verbalise why.

So what does thinking style have to do with writing?  Understanding how you and your co-authors think may help you in planning your paper. Visual thinkers may like to plan their paper with flow charts, colours, and arrows, to later turn these images into words \cite{Ashby2005}.  On the other hand, verbal thinkers may prefer making a classical outline or using bullet points to plan their paper.  Understanding that your colleagues may have a different thinking style may also help avoid conflicts in writing.  If for example your supervisor says you must begin by making an outline {using text or bullet points}, but that does not work for you, then explain that you work better with coloured flow charts or perhaps make your flow chart and use that as a step in developing an outline that your supervisor and co-authors can understand.

\subsection{Ten Steps to Plan Your Research Paper}

Construct your paper carefully, as if producing a painting by starting out with a rough sketch and then filling in the details. Do not start off by sitting down with a blank sheet of paper or newly opened file, and pretend you will simply start writing at the beginning and keep writing until the end. Even the best writers find this almost impossible to do. 

The following would be our idea of planning and writing a paper in ten steps, from start to submission:

\begin{itemize}

\item First, define the message you want to convey to the reader, and what story you want to tell. This step in particular may require the kind of critical thinking and literature review discussed in Sect. \ref{sec_think} because readers will ask those same questions when reviewing your paper. Therefore, define your message clearly, and writing will become possible. Fail to do so and you will struggle. 
\item Second, draft the title and abstract of your paper (discussed in Paper II).
\item Third, brainstorm and sketch out how your paper will look. Your personal preference will steer you towards creating an outline with words, pictures, or colours (Sect. \ref{subsec:think_style}).
\item Fourth, define your sections (typically: introduction, methods, results and discussion) and subsections and then add the key figures and tables that you want to include in the right place. Draft the captions. We discuss the details on how to write these sections in Paper II.
\item Fifth, add paragraphs, identified by their contents and what you need to say (Paper II).
\item Sixth, fill in the paragraphs and (sub)sections with a rough first draft of the text. You may want to start with the easiest sections (usually sample, data, methods) and leave the hardest for later (discussion).
\item Seventh, cite the sources of ideas that are not your own or paraphrase them to not forget about them later. Many journals automatically scan uploaded manuscripts for plagiarism (Sect. \ref{plagarism}) so we recommend keeping track of your references as you draft your paper. If you are quoting a definition or standard description, put the quoted text between quotation marks and provide the source. If you use any of the material discussed in this series, you may want to cite and/or acknowledge Paper I and II. 
\item Eighth, edit your paper. We cannot emphasise this point enough. Anything that does not contribute towards the message of your paper should be cut out (or consigned to the appendix). Do not be afraid to delete -- not every single paper you read or piece of work you did may be worthy of inclusion.
\item Ninth, more and more and more editing and polishing (including incorporating comments from others). It is useful to take a break from writing your paper. Do not work on it for a couple of weeks, come back to it with fresh eyes and edit some more.
\item Last and extremely important step: decide that enough is enough, or run up against a deadline, and submit your paper to the journal of your choice (Sect. \ref{sec3}). {\it Usually a deadline rather than having reached perfection determines when your paper is finished.}

\end{itemize}

If you are not sure about a fact or figure, do not let this slow down your writing. Just make a note to yourself to add the missing bit of information later. Always ask others to read your work before you submit it. Apart from your co-authors, you can ask supervisors, or colleagues, or friends, or family members to read your draft. It can be a good idea to share early drafts for continuous feedback, instead of stressing yourself for a perfect near-final version. This allows ample time for you to get help with or focus on issues that you are unsure about (e.g. spelling, grammar, mathematics or literature completeness). \par

\subsubsection{\LaTeX}

As you start out as an author for astronomy or astrophysics journals, it is worth investing time and effort to master at least the basics of \LaTeX. \LaTeX\ is often used to prepare manuscripts in astronomy and in much of physics and engineering. It is a software system for document preparation which offers significant advantages over word processors such as LibreOffice Writer, Microsoft Word, or Apple Pages. Advantages of \LaTeX\ are that the software is free, the document is prepared in plain text, and that it is excellent at typesetting mathematical expressions as well as non-Latin scripts. \par 
\LaTeX\ is also a tool that can be paired with reference managers. Writers should never type references by hand because such an exercise leads to a source of error in the literature. There are several reference managers you may choose from (e.g. \href{https://endnote.com}{EndNote}), but even starting out with a simple .bib file is a great first step. \par 

Many text editors and software packages that can compile a \LaTeX\ document are freely available for all main computing platforms. For instance, \href{http://www.overleaf.com/}{Overleaf} is increasingly used as an online tool to allow multiple co-authors to share the editing of a \LaTeX\ document. Additionally, most journals provide their own \LaTeX\ macro packages to prepare manuscripts. Therefore, as you start putting your paper together, if you use the \LaTeX\ macro package of your favourite journal (more on this in Sect. \ref{sec3}), your manuscript will immediately start to look `real'!

\subsubsection{Data and Software}

A key aspect of ethics in science is that scientists are open about 
their work, what they have done, and about the data and tools they 
used. Whenever you can, publish your data and/or your software or 
code with your paper. Both \href{https://academic.oup.com/mnras}{MNRAS} and \href{https://www.aanda.org}{A\&A} have a contract with the 
\href{https://cdsweb.u-strasbg.fr/}{CDS} which guarantees long-term 
archival. \href{https://git-scm.com}{Git} or 
\href{https://github.com}{GitHub} are often used to publish software. However, as they are not permanent repositories, one should instead use \href{https://figshare.com}{Figshare}, \href{https://datadryad.org/stash}{Dryad} or \href{https://zenodo.org}{Zenodo}. A useful list of repositories can be found in this \href{https://www.nature.com/sdata/policies/repositories}{link}.

When you publish your software or code, ideally you license it under a 
free-software 
\href{http://www.gnu.org/philosophy/free-sw.html.en}{license}. If you 
do, you grant your colleagues the right to `run, copy, distribute, 
change and improve' the software. It is interesting that if you make 
your software publicly available but without an explicit licence, it
is copyright-protected under the Berne Convention Implementation Act of 
1988 and other
astronomers are legally forbidden from copying the software, modifying 
it, distributing it, or distributing a corrected or improved version (as 
pointed out in the Ethics 
\href{https://eas.unige.ch//documents/EAS_Ethics_Statement.pdf}{Statement}
of the European Astronomical Society).

Avoid making data available through your own website, or your 
institute's, or your research project's. The websites will disappear 
with time, or the URLs will change. The only way to guarantee that your 
data or software will remain accessible (even after you have changed 
your career path, or have retired) is to formally publish it in a 
repository. We recommend \cite{Chen2022} for an excellent discussion on 
how to publish data in astronomy.

\subsection{Avoiding Plagiarism}
\label{plagarism}

Avoiding plagiarism while preparing your paper is so important that it deserves a section by itself. To plagiarise is to copy, use, distribute or present someone else's intellectual property (text, results, ideas, etc.) as your own and/or without proper acknowledgement. All professional research institutions take disciplinary action against plagiarism so make yourself familiar with these policies.  \par 
Plagiarism may arise for several reasons.  For example, you cannot think of a better way to express an idea or concept in your own words compared to the original author.  Perhaps you copy and paste
a text from a publication, thinking `I will change it later' only you forget to change it before you finalise your draft, and end up plagiarising. To solve these problems, here are a few guidelines: 

\begin{itemize}
\item Improve your vocabulary and check sources like the Manchester Academic {\href{https://www.phrasebank.manchester.ac.uk/}{Phrasebank}} for ideas to expand your own vocabulary of words and phrases. However, keep in mind that academic writing does not always mean to use the most fancy, flowery or sophisticated words. Aim for clarity and simplicity.
\item Phrases such as `As argued/discussed/suggested by Author et al. (year) ...' or `Author et al. (year) point out that..' are common ways in which you can immediately inform readers that the idea or discovery which follows do not belong to you. 
\item Do not take notes from other publications by copying and pasting statements---including those from your past publications (`self-plagiarism')---always use your own words or rephrase statements. Take the time you need to think deeply about the concept or idea you read and then want to write about (Sect. \ref{sec_think}). You can always find the right 
words to express your thoughts in a thesaurus or on the web (DB frequently uses a thesaurus, while NC and JHK are fans of typing `$<$word$>$ 
synonym' into google).

\item Do not save on citations---whether you use a direct quote from a publication or not, cite the necessary publications that you used to explain or provide context to the subject of your paper. However, do check that a given publication indeed supports the statement that you claim it does!

\end{itemize}

A full discussion of these and other tips on avoiding plagiarism is in 
\cite{Roig2015}.

\subsection{Basic Points on Writing Style}

A refereed research paper is written differently from an outreach article, grant proposal or job application. In general, the basic points to consider when you write academically or non-academically are as follows:

\begin{itemize}
    \item {\bf Audience}---who are my readers, what is their level of expertise and what do they need to know?
    \item {\bf Purpose}---is my aim to inform (science paper) or entertain (popular)?
    \item {\bf Tone}---should my piece be objective, neutral or emotional?
\end{itemize}
    
Given the aims of this paper, we answer the above questions specifically for scientific research papers in refereed journals. Firstly, your audience will be professionals in the field. However, they will not necessarily be experts in your sub-field, or aware of all the methods, techniques or tools you use. Therefore, you will need to explain all the specifics needed to understand your work, while being careful to avoid jargon or explaining the obvious.

Secondly, the purpose of a scientific paper is to inform, to describe your experiment or theory and results in a scientific way. In other words, that sufficient detail and references are presented in your paper such that your results and conclusions can be verified and/or reproduced.  For example, avoid statements such as `we reduced the data in the standard way' but rather mention `we followed the standard procedures for data reduction as described in detail by Author et al. (year)' or `we reduced the data in the standard way by first ..., then ... and finally ... .'\par 

Finally, the tone to be used in your research paper is formal. For instance, do not use contractions (use do not, cannot, will not, it is, etc., rather than don't, can't, won't, it's), do not address the reader directly in the imperative (say `the data are...' and not `note that the data are...'), and avoid colloquialisms and slang, or references to fashionable popular culture.  \par 

A formal tone also implies that any criticism of others' work should be phrased in professional terms, polite, and without personal sneers. Instead of the insulting `X et al. clearly showed their lack of understanding of the basics of statistics when they published...' one could be highly critical but still professional: `The recent study of X et al. fails to properly account for ... in their statistical analysis.'. Writing your research paper in an open, honest, respectful and professional tone is crucial to create an inclusive environment for all scientists, even if you disagree with previous works.\par

\subsubsection{Towards Inclusive Language}

In relation to the tone of a research paper, in recent years the field of galaxy formation and evolution in particular has seen the emergence of a number of terms which in real life signify violent, highly unpleasant and mostly illegal acts, sometimes with misogynous or racist undertones. Some of these terms, such as a hierarchical scenario, cannibalism, stripping, strangulation or starvation, are by now so integrated into the professional vocabulary that in particular younger scientists may consider their use `normal'. We would urge exercising restraint in the use of violent terminology, and to consider using alternatives whenever possible. More collaborative and inclusive terms may be used to replace the violent ones, including an ancestral scenario, collectivism, sharing, collaboration and preservation for the five mentioned above as examples \citep[see][]{Vallejo2018}. You can define these terms when you first use them in the text to ensure that readers understand what you mean. Alternatively, you may describe the main idea behind these terms using your own words and cite the authors who first proposed the idea as the reference throughout the paper, e.g. `As proposed by Author X et. al...' and later on in the paper `The theory described by Author X predicts...' \par 

\section{Choice of Journal and the Road to Publishing}\label{sec3}

How you need to craft your research paper as discussed in Sect. \ref{sec2} may also depend on which journal you choose to publish your work in. In this section, we discuss the choice of journal for submission and the publication process for your research paper.

\subsection{Where to Submit?}

You may be invited to submit a research paper, for instance to include in the proceedings of a conference you have attended, or to contribute an invited paper. If you are not invited, however, you and your co-authors will need to decide where to submit your manuscript. Some general guidelines to help you are as follows:

\begin{itemize}
\item Consider the field, scope and readership of the journal, and check the author's instructions to make sure your paper `fits'.
\item For your career development, refereed professional journals are best. 
\item Aim as high as is reasonable. Impact is important: a `famous' or quartile-1 (Q1) journal is better than a less-known one (the \href{https://www.scimagojr.com/journalrank.php?category=3103}{SJR} Journal Rankings will give you a good idea).
\item Do not decide {\it a priori} that a certain journal will not accept your paper. Let the editor and referee arrive at such a decision. Ask advice from more experienced researchers before submitting your paper if you are not sure.
\item Well-reputed free-to-publish journals exist but several also ask you to pay page charges. You can request that they be waived if you have no funds. Do ask before submitting your manuscript, not after it has been refereed!
\item There are many other refereed journals but most do not have the same impact as the more famous journals and may not make the same impression on your curriculum vitae. Many are commercial so check any costs before agreeing to publish. Some journals are even considered `predatory' \cite{Eriksson2017}. These journals may have low scientific standing, but by the use of clever titles that sound like real journals, they may aggressively pursue your submissions in order to collect publication fees.
\item Once you have made a choice, follow the author's instructions to produce your manuscript for submission. These 
will include instructions on style or length, but also on whether to use 
American or British English, which have differences in spelling (for 
example, the American -or words are written with -our in British 
English:  color/colour, neighbor/neighbour; American English ends some 
words with -er, while British English uses -re: center/centre, 
meter/metre; American English has words that must end in -ize or -yze, 
while British English also writes them with s: analyze/analyse, 
categorize/categorise).
\item If a paper is rejected, analyse why (ask your co-authors or other colleagues with more experience) and if you do not agree with the decision, consider submitting a revised version of your paper to the same journal, or to another journal.

\end{itemize}

\subsection{Authorship Considerations}

Single-author papers are rare in modern astronomical research. The first author is often the person who has done most of the work, by some integrated measurement. In large collaborations, author lists are sometimes alphabetical and the person who can claim main authorship may in that case be identified as the `corresponding author'. In many fields the last position in an author list is an honorary one, often given to a key senior colleague. In astronomy, however, this person might be second or third in the list. As evaluations for jobs, promotions, grant competitions, etc. normally consider first positions on an author list (and second and third to some extent), it is important that you explain what your role has been if you were not first (or 2nd, 3rd) author---and that you ask your referees to confirm this in their letters.

Co-authors are typically those who made substantial contributions to a paper. They can include colleagues who had the original idea or developed the instrumentation or code, or put the funding in place. People who only were on, e.g., the observing proposal or provided limited assistance during an observing run typically are acknowledged rather than made co-author. Co-authorships can be offered to students by researchers outside their supervisory team when the student has contributed to the paper, or forms part of a larger collaboration. 

There are no universally accepted rules for who will become a co-author but you may find authorship \href{https://www.nature.com/nature-portfolio/editorial-policies/authorship}{policies} useful when writing your own paper. Our advice is that as a junior scientist you should not unilaterally exclude prospective co-authors unless you particularly fancy interpersonal conflict. If you have doubts, ask a senior colleague for advice. Many papers have co-authors who have done little, and who probably should merely have been acknowledged. On the other hand, people may have made key contributions to a project at an earlier stage, which are not visible to latecomers but remain vital. As it is such a grey area and people may hold strong views, it is perhaps best to accept as a fact of life that every now and then a person becomes your co-author who you would personally not have given that much glory. 

\subsection{On Names and Numbers}

It is a good idea to consider how you want to be known in the scientific world before you submit your first paper. Names are much affected by local culture. For instance, Chinese nationals and many others start their name with what in English would be the last name, followed by their first (or given, or fore-) name(s). Some female scientists prefer to sign with initials rather than first names so they are not immediately identified as female (current authors NC and DB do not). In some regions, people only have one name (they are {\it mononymous}) and not all editors and publishers are ready to accept this.

You should consider the conventions in our field along with your own preference when choosing how you will be named in scientific publications. Citations are key, and it is important that your papers be readily identified with you. Decide which name(s) to use and which additional initials or other signs to maintain. If the first name you are known by does not correspond with your initials, be careful and at least consistent in how you publish your name. Spanish speakers often hyphenate their two last names to avoid being cited with their second last name only (the first last name being considered a middle name, or second first name!). Some journals now allow authors to add their name in Chinese, Japanese or Korean characters after the English version of their name. Another positive novelty is that several journals now allow authors to change their names (for reasons including gender identity, marriage, divorce, or change of religion). 

We recommend that all authors register with {\href{https://orcid.org/}{ORCID}} to obtain a free persistent digital identifier (ORCID iD) which is basically a number that is tied to your person. By coupling your professional information, papers but also affiliations, grants, etc., with your ORCID iD, all your products can be identified as yours even if searching by your name does not yield unique, conclusive or complete results. 

\subsection{Roles of Referee, Editor, and Copy Editor}

You write your paper, typeset it to perfection using \LaTeX, and submit it to the journal. A scientific editor will consider your manuscript and decide whether it might be suitable for publication in their journal. An editor is usually a senior scientist and a subject specialist. They may send your manuscript to a referee (in astronomy, usually only one, in related fields, up to five) for peer review. The referee is a colleague who is a specialist in the topic of the paper. They will write a report containing a recommendation to the editor and suggestions for improvement to the authors. The role of a referee is to make sure the paper is scientifically sound, but the referee is not a co-author of the paper. A referee can thus recommend acceptance without agreeing with everything written in a paper, or with how it is written. \par 

In astronomy, most referees are supportive, and after authors submit a revised version of their manuscript, taking into account the recommendations of the referee, most papers are accepted for publication. Sometimes multiple rounds of refereeing and revision are needed or a second referee is sought to adjudicate in a stalemate situation. Other times a paper is rejected. When dealing with difficult referee reports, is it always a good idea to consult with more experienced colleagues. More background information on the roles of editors and referees at the journal A\&A is provided in \cite{2004A&A...420E...1B}. \par 

Once a referee recommends acceptance and the editor indeed accepts your paper, it will enter the production stage. A copy editor will typically proofread your manuscript and make changes to perfect both English language usage and compliance with the journal or publisher house style. Editors are usually very good at their job, but they are not normally astronomers. So if you are asked to check the page proofs (and answer any queries a editor may have identified) it is very important to check your paper carefully to make sure no inadvertent changes have been made. As the editor's changes are usually identified, this is also a learning opportunity to see where your phrasing or typesetting was not optimal. A\&A has an instructive {\href{https://www.aanda.org/contacts-bottommenu-162/69-author-information/language-editing}{list}} of things that their editors often need to correct, which is well worth looking through. 

At the page proof stage you can still make small changes or additions if absolutely necessary (such as including a missing reference that has been pointed out to you after you pre-published your paper on the preprint server \href{https://arxiv.org}{ArXiV}). After this step, the paper is typeset, gets a formal journal reference, \href{https://www.doi.org}{DOI} and is published. It is indexed and will form part of the body of scientific literature---in perpetuity!

\section{Conclusion}\label{sec4}

In this first part of our guide to scientific writing for astronomers (Paper I), we have provided a discussion on how to plan your research paper in astronomy, including defining  the main message, outlining the paper, and adapting the style to meet the needs of the audience. We also emphasise the importance of organising your research paper following the standard practices and guidelines provided by your choice of journal. The second part of this guide (Paper II) will cover how to organise and write each section of a research paper, based on the typical IMRaD (=Introduction, Methods, Results and Discussion) format. Paper II will also deal with the mechanics of writing (i.e. paragraph structure, sentence linkage etc.), especially geared towards non-native speakers of the English language. \par 
Finally, to our readers, we emphasise that writing is a process that can be learned and constantly improved. Many of our recommendations can take several years to truly master but we hope this guide provides a clear framework to begin planning and writing your research paper in astronomy/astrophysics or serve as a reference throughout your scientific writing career. 

\backmatter

\bmhead{Acknowledgments}

We thank S\'ebastien Comer\'on, Sim\'on D\'\i az-Garc\'\i a, Erik \& Laura Knapen Almeida, Cristina Mart\'\i nez-Lombilla, Rainer Sch\"odel and Aaron Watkins for comments on an earlier version of these notes. NC thanks T. Emil Rivera-Thorsen, Matteo Pompermaier and Chris Usher for interesting discussions. Part of this paper is based on a scientific writing course delivered by J.H.K to mostly MSc and PhD students in Ethiopia and Rwanda. He wishes to thank Professors Mirjana Povi\'c and Pheneas Nkundabakura for organising those courses, and the students for participating.

\section*{Declarations}

\bmhead{Funding}

J.H.K. acknowledges financial support from the State Research Agency (AEI-MCINN) of the Spanish Ministry of Science and Innovation under the grant `The structure and evolution of galaxies and their central regions' with reference PID2019-105602GB-I00/10.13039/501100011033, from the ACIISI, Consejer\'{i}a de Econom\'{i}a, Conocimiento y Empleo del Gobierno de Canarias and the European Regional Development Fund (ERDF) under grant with reference PROID2021010044, and from IAC project P/300724, financed by the Ministry of Science and Innovation, through the State Budget and by the Canary Islands Department of Economy, Knowledge and Employment, through the Regional Budget of the Autonomous Community. NC acknowledges support from the research project grant `Understanding the Dynamic Universe' funded by the Knut and Alice Wallenberg Foundation under Dnr KAW 2018.0067.

\bmhead{Authors' contributions} 
 
The idea of this series stems from a \href{https://www.astro.rug.nl/~sundial/blog/index.php/2019/10/28/a-workshop-on-scientific-writing-in-tenerife/}{workshop} by the name `Scientific Writing for Astronomers', organised by N.C. and D.B. 
at the Instituto de Astrof\'isica de Canarias, Tenerife, Spain in October 2019. All authors further developed the ideas presented in this manuscript. 
N.C. and J.H.K. are the primary authors of all sections. D.B. wrote the first draft of sections 2.2 and 2.4, and was involved in editing the final manuscript.

\bmhead{Competing Interests} 

The authors declare no competing interests.

\bibliography{NatAst_P1}% common bib file
%% if required, the content of .bbl file can be included here once bbl is generated
%%\input sn-article.bbl

%% Default %%
%%\input sn-sample-bib.tex%

\end{document}